\def\year{2022}\relax
\documentclass[letterpaper]{article} 
\usepackage{aaai22}  
\usepackage{times}  
\usepackage{helvet}  
\usepackage{courier}  
\usepackage[hyphens]{url}  
\usepackage{graphicx} 
\usepackage{natbib}  
\usepackage{caption} 
\DeclareCaptionStyle{ruled}{labelfont=normalfont,labelsep=colon,strut=off} 
\usepackage{algorithm}
\usepackage{algorithmic}

\usepackage{newfloat}
\usepackage{listings}
\floatstyle{ruled}
\newfloat{listing}{tb}{lst}{}
\floatname{listing}{Listing}

\usepackage{xcolor}
\newcommand{\answerYes}[1]{\textcolor{blue}{#1}} 
\newcommand{\answerNo}[1]{\textcolor{teal}{#1}} 
\newcommand{\answerNA}[1]{\textcolor{gray}{#1}} 
 
\title{Author Unknown: Evaluating Performance of Author Extraction Libraries on Global Online News Articles}
\author {
    Sriharsha Hatwar\textsuperscript{\rm 1}, 
    Virginia Partridge\textsuperscript{\rm 1, 2}, 
    Rahul Bhargava\textsuperscript{\rm 3, 2}, 
    Fernando Bermejo\textsuperscript{\rm 2}
}
\affiliations {
    \textsuperscript{\rm 1} University of Massachusetts, Amherst\\
    \textsuperscript{\rm 2} Media Cloud\\
    \textsuperscript{\rm 3} Northeastern University\\
    shatwar@umass.edu, vcpartridge@umass.edu, r.bhargava@northeastern.edu, fernando@mediacloud.org  
}

\begin{document}

\maketitle

\begin{abstract}
Analysis of large corpora of online news content requires robust validation of underlying metadata extraction methodologies. Identifying the author of a given web-based news article is one example that enables various types of research questions. While numerous solutions for off-the-shelf author extraction exist, there is little work comparing performance (especially in multilingual settings). In this paper we present a manually coded cross-lingual dataset of authors of online news articles and use it to evaluate the performance of five existing software packages and one customized model. Our evaluation shows evidence for Go-readability and Trafilatura as the most consistent solutions for author extraction, but we find all packages produce highly variable results across languages. These findings are relevant for researchers wishing to utilize author data in their analysis pipelines, primarily indicating that further validation for specific languages and geographies is required to rely on results.
\end{abstract}

\section{Introduction}
Online news articles are a valuable source of information for understanding unfolding events and the overall social environment, but as unstructured data they can be very hard to track and investigate through computational means. Digital analysis of online content is now a common approach taken when studying the production and dissemination of news.  Researchers across various fields use algorithms and heuristics on small- and large-scale corpora of HTML-based news articles to enable meaningful exploration of topics such as harassment and hate speech detection \cite{zannettou_measuring_2020}, event detection \cite{piskorski_online_2011}, media polarization \cite{xu_what_2020}, the spread of misinformation \cite{jiang_traditional_2017,sarr_factextract_2018}, partisan framing \cite{chen_partisan_2023}, and beyond. Advancements in Machine Learning (ML) and Natural Language Processing (NLP) have paved the way to extract structured content from corpora of news articles \cite{banos_scalable_2015, roberts_media_2021}, allowing for more complex computational analysis. However, using these approaches in research settings relies on robust validation of the underlying content extraction techniques. Leveraging well-performing methods facilitates the transition from unstructured HTML text to distinct properties, including but not limited to the publication date, publication source, primary image, primary language, article title, article content, and author information. The final property, authorship, proves particularly valuable for investigating issues such as gender equity \cite{ryskina_gender_2022, boczek_gender_2023, mitchelstein_whose_2020}, news consumer behavior \cite{goyanes_effects_2021}, identification of dominant voices \cite{tanner_authorship_2011, armstrong_influence_2004}, monitoring the dissemination of narrative frames \cite{tsagkias_linking_2011, kian_framing_2009}, understanding evolving media ecosystems \cite{quandt_no_2008}, and other subjects.

Multiple off-the-shelf solutions exist for researchers wishing to run author extraction on a corpus of online news articles. In the context of this paper, we use “author extraction” to mean the automated processing of HTML news content to discern the individual(s) or organization(s) identified as the author(s) of the text content, utilizing the mix of structured and unstructured data source material found on the web. This can be further specified as “byline extraction” when working in the context of online news articles published by media sources, but we will use the more generic “author extraction” for purposes of our discussion. In cases where news articles already include author information in their structured, machine-readable HTML metadata, the extraction process performed by these libraries is straightforward (though it might still involve conflict resolution if multiple types of content indicate different authors). However, in other instances, extraction necessitates treating HTML as semi-structured or unstructured text. The methods existing libraries employ can broadly be grouped into three categories:
\begin{enumerate}
\item ML methods based on hand-labeled datasets;
\item Heuristic methods that rely on a webpage's HTML structure and metadata tags;
\item A combination of ML (1) and heuristic methods (2). 
\end{enumerate}

Despite widespread use of existing author extraction libraries in computational journalism, media communications, journalism studies, and beyond, there is a scarcity of systematic reviews evaluating their performance. This is particularly true for global, non-English language contexts and corpora that include more contemporary web coding norms. To address the gap in validation work, this paper contributes a cross-library evaluation of the performance of five extant software packages that perform author extraction, along with a baseline transformer model, against a new hand-coded cross-lingual dataset (covering 715 authors of 754 online news articles across 10 languages). This offers both a new benchmark dataset and a point-in-time analysis of library performance, supporting more robust research applications using author extraction libraries and also the development of performance improvements that would benefit many fields of research that utilize author extraction.

\section{Literature Review}
This paper builds on work in three distinct domains. First, we pull motivation from the widespread use of author information in a broad set of research related to online news information. Second, we follow practices of participatory benchmark development for computational tasks related to extraction of structured information from unstructured content. Third, we evaluate implementations of a variety of algorithmic extraction techniques. Each is summarized in more depth in this section.

\subsection{Motivations and Use for Author Data}
Authorship is a key element of analysis of digital text in general, and of news production in particular. Researchers active in journalism studies, mass communication, NLP, and other fields utilize author information to support their broader work on news and publishing. In journalism and communications there is a significant body of work on identifying the gender of authors to evaluate both bias \cite{boczek_gender_2023} and impacts on story content \cite{armstrong_influence_2004}. Author information is even used to measure impacts on news readers’ desire to pay for content \cite{goyanes_effects_2021}. Communication scholars use author information to assess the frequency of using newswire reporting vs. original reporting \cite{shaheen_over_2019, tanner_authorship_2011}, and also to investigate connections between authorship and biases in content presentation and  source selection for news reporting \cite{armstrong_influence_2004, somaini_who_2018}. Web scholars utilize author information as part of larger investigations into the spread of stories through social networks \cite{tsagkias_linking_2011}. Both the unbundling of traditional news products (such as newspapers) and the possibility of automating the  production of news are making authorship information even more relevant to understanding the news ecosystem \cite{waddell_attribution_2019, montal_i_2017}. 

Others working in academic scholarship use author information to measure and track disparities in publishing \cite{ryskina_gender_2022, vogel_he_2012}, to analyze research outputs  \cite{rosen-zvi_author-topic_2004}, and to build networks of collaboration and affinity in scientific production \cite{do_extracting_2013, bihari_key_2016, rodighiero_mapping_2018}. But the challenges posed by authorship extraction in news content has hindered the analysis of these networks in the world of journalism, where researchers have had to rely for the most part on social media (Twitter / X) data to identify connections and communities \cite{li_journalists_2023, vergeer_peers_2015}. 

\subsection{Benchmarking Metadata Extraction from Online News}

Automated extraction of structured data from webpages is a well-established field, typically involving the collection of appropriate web content and then the development of a manually-created benchmark to evaluate techniques against. One such notable dataset is SWDE \cite{hao_one_2011}. It contains eight subject areas/verticals, each of which contains ten sites with a total of 120,491 web pages (4k-20k pages in each subject/vertical). For example, the vertical containing web pages about books has around 20k web pages and contains information about the title, author, ISBN-13, publisher, and publication date. This dataset has all the HTML components for a particular webpage and does not contain any images associated with the webpage. Similarly, the Klarna product page dataset contains a rich resource for web scraping and visual recognition tasks in ecommerce \cite{hotti_graph_2021}. It features 51,701 diverse product pages from 8,175 websites across 6 European languages (English, German, Finnish, Swedish, Norwegian, and Dutch). Each page is  labeled with key information like price, name, and main image, allowing for accurate extraction and analysis. The inclusion of both HTML code and screenshots helps in the development of robust algorithms that can handle the complexities of webpage rendering. 

In the domain of online news content, Varlamov et al. designed a benchmark dataset that contained news webpage articles and manually annotated metadata such as publication date, title, etc., all collected from Russian language media sites (724 web pages from 114 news websites)  \cite{varlamov_dataset_2022}. Methodologically, they used LabelStudio for the markup process, annotating textual spans in each webpage. They also evaluated the extraction of these attributes using several online tools, LSTM, and transformer based models. Barbesi et al.  evaluated the extraction of the main text contents of a webpage (the part displayed centrally without left or right bars) with existing tools published under open-source licenses \cite{barbaresi_out---box_2020}. The dataset used for the task was DAnIEL, a multilingual human-annotated dataset originally designed for tracking online news about epidemics. DAnIEL consists of news articles in their original HTML format in Chinese, English, Greek, Polish, and Russian.

\subsection{Computational Approaches to Author Extraction}
Recently, there have been many approaches using rule-based heuristics, ML models or a combination of these to tackle the problem of author extraction from online web articles. One of the most well-known heuristic tools is Trafilatura\footnote{\url{https://trafilatura.readthedocs.io}}. Combining existing content extraction libraries to parse HTML with fallbacks that account for many content management systems, this general purpose web-scraping library is capable of extracting a variety of elements from DOMs, including authors \cite{barbaresi_trafilatura_2021}. Other open source libraries such as Newspaper3k\footnote{\url{https://newspaper.readthedocs.io/}} and Readability\footnote{\url{https://github.com/mozilla/readability}} similarly combine regular expressions, HTML parsing and cascading rules to extract metadata from webpages.  

Common ML approaches to author extraction in the literature rely on extracting textual, visual or spatial relationships from manually annotated webpages and using these features to train a support vector machine (SVM) or decision tree classifier. In one of the earlier works, Changuel et al. focus on automatic author extraction from HTML articles especially when there is no meta $<$author$>$ tag.  Assuming there is only one author per web page, they constructed a training and evaluation dataset of 354 articles using a combination of manual annotation with semi-automatic article extraction using gazetteers and web search APIs.  Their algorithm initially extracts person names using a DB of US first names list from a DOM tree, constructs a context window of interests as features that indicate information about, among other elements, the position of the name and the possibility of data in its vicinity. Each person's name on a web page is a potential author, and the final extraction task is formulated as a binary classification problem with names and corresponding features  fed to a decision tree model which achieved an  80\% F1 score \cite{changuel_automatic_2009}.  Kato et al. took a ranking approach to identifying the author or agency responsible for a web page in Japanese, using linguistic features like part of speech to identify potential authors then ordering these candidates with a Ranking SVM using local features like their proximity to the main content by flattening the DOM tree of the webpages \cite{kato_extracting_2008}. Do et al. present a unique way for extracting authors and their affiliation in scholarly articles, solving this task using a two step process. The first step involves converting a PDF to XML format with optical character recognition  and feeding it to a Conditional Random field model to extract the author and affiliation present in the article. Later, the extracted text is processed and fed to a relational classifier such as SVM to associate the author to the affiliation. This approach achieved high F1 measure on a manually annotated dataset of scholarly articles across many disciplines \cite{do_extracting_2013}.

Other approaches side-stepped the expensive process of annotating author data by constructing training datasets via unsupervised methods. For example, using the insight that users over select unique strings for profile names, Liu et al used n-gram probabilities to identify pages with unique author names, building a dataset for author extraction from user generated content pages.   With a variety of visual, text and content features as input, training a SVM based classifier with RBF kernel on this dataset was found to perform better on author extraction than a model trained on supervised data \cite{liu_unsupervised_2012}.

\section{Methods}
The core processes for this study involved creating a benchmark dataset to evaluate against, selecting oft-used tools to evaluate, and developing criteria for comparative analysis of results.

\begin{figure*}[t]
\centering
\includegraphics[width=0.8\textwidth]{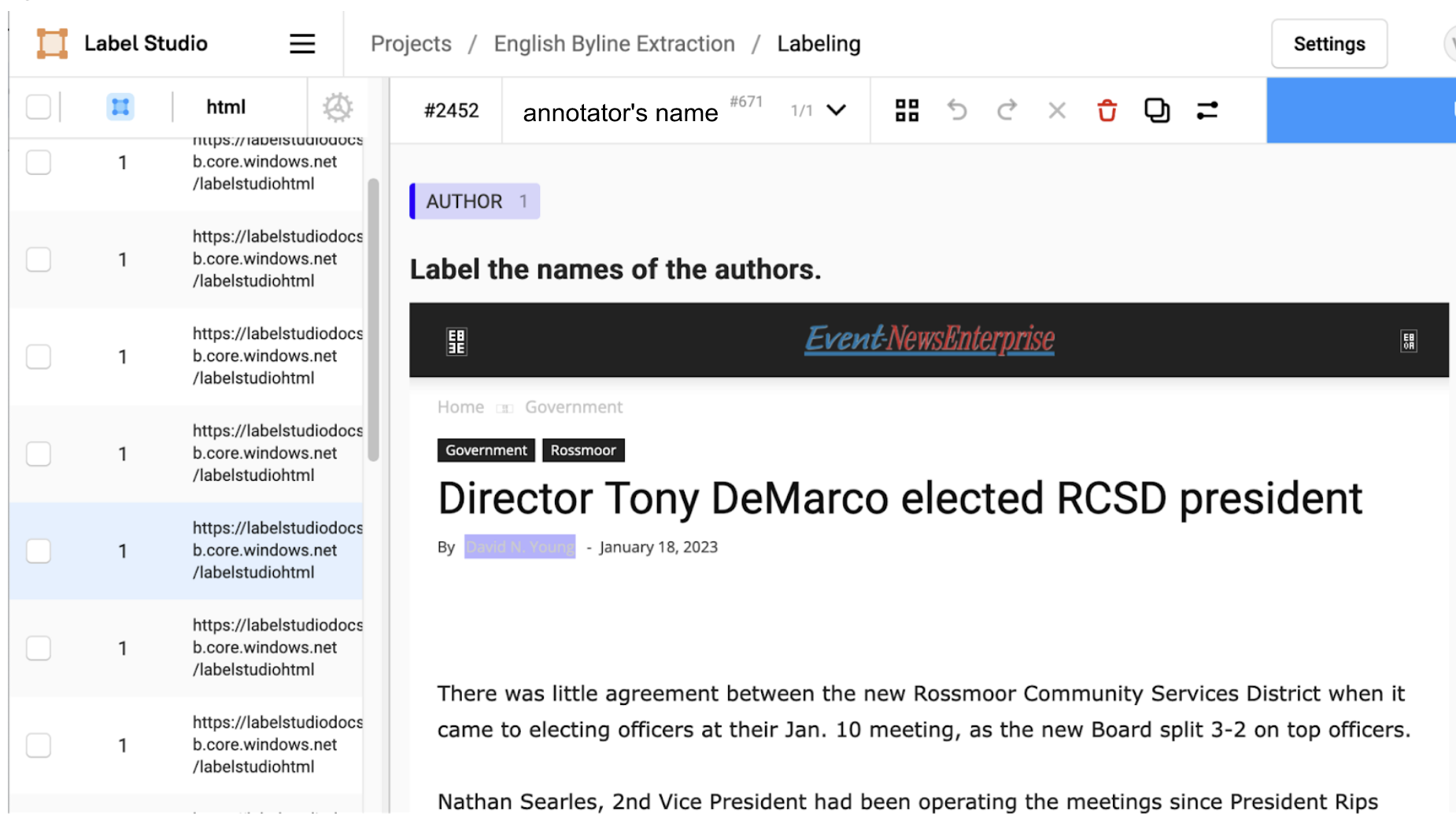} 
\caption{Annotation interface in LabelStudio showing an author annotated in an HTML document as it would appear on the original website.}
\label{fig:labelstudio_ui}
\end{figure*}

\begin{table}[t]
\centering
\begin{tabular}{|p{0.25\linewidth}|p{0.25\linewidth}|p{0.2\linewidth}|}\hline
    \textbf{Language, \mbox{2-letter code}} & \textbf{Total \mbox{Documents}} & \textbf{Number of Authors} \\ \hline
    Danish, da  & 30 & 54 \\ 
    German, de & 99 & 91 \\
    Greek, el & 32 & 23 \\
    English, en & 94 & 99 \\
    Spanish, es & 30 & 24 \\
    French, fr & 100 & 81 \\
    Hindi, hi & 100 & 41 \\
    Russian, ru & 94 & 125 \\
    Urdu, ur & 100 & 92 \\
    Chinese, zh & 75 & 85 \\ \hline
\end{tabular}
\caption{Counts of documents and number of author annotations total in the final dataset.}
\label{table:dataset_stats}
\end{table}

\subsection{Production of Evaluation Data}
Data collection and annotation for evaluating author extraction across different linguistic and cultural settings poses several challenges. The annotation scheme must account for well-known variation, such as articles having multiple authors, anonymous authors, or attribution to an organizational author (as is often seen with newswire services like Reuters). The scheme must also be flexible enough to capture variation we could not foresee when designing the task, due to cultural or linguistic biases. Additionally, we needed to respect the valuable and limited time and effort of annotators with the linguistic skills and familiarity with communication and journalism studies necessary to complete this task. 

Our partner in this endeavor was \url{anonymizedURL}, an open-source media research project that works with partners around the world to study the flow of news and information. We recruited and surveyed their research and development partners, asking them to self-report which languages they read well enough to assist with an author/byline labeling task. Based on those responses, we collected articles in each language from the organization's online news archive, randomly selecting ten days between January 1st, 2023 and April 30th, 2023, then pulling ten articles from each day for each language, totalling 100 random articles for each language. Separate annotation tasks were configured for each language in LabelStudio, where annotators could easily log in, select their desired language, view articles in the same way they appeared on the original website, and highlight author information, as seen in Figure \ref{fig:labelstudio_ui}. Although we started with 18 languages, we did not ultimately have enough resources to annotate all articles in every language; the final dataset includes the 10 languages where annotators were able to annotate at least 30 articles, as seen in table  \ref{table:dataset_stats}. This set of ten languages covers a variety of language families, orthographic scripts, and geographic locations.

To develop a codebook we began with an approach of grounded theory \cite{blandford_qualitative_2016}. The initial round of annotations were performed by close collaborators, including a media librarian, social data researcher, and one of this paper's authors. The key task prompt was to highlight ``the part of the page that introduces the author or serves the purpose of attributing the content to them." Common patterns identified by this core pilot group were then formalized by the research team into guidelines that were distributed with the rest of the articles to all annotators. Although annotators were encouraged to follow the guidelines whenever possible, we asked them to alert us to examples of articles that did not conform to these guidelines or were difficult to annotate. The broader team of annotators included digital archivists, creators of existing author extraction libraries, media research scholars, and engaged news readers. As a final step in preparing the dataset, we removed any articles which annotators could not mark due to issues with language identification or formatting, such as not being in the expected language, having author names in images or mangled formatting in the rare cases where an article appeared differently on LabelStudio than the original website. Articles that did not explicitly indicate an author or attribute credit were kept in the final evaluation dataset in order to catch false positives. The final dataset statistics are given in table \ref{table:dataset_stats}. There were not enough linguistic resources to have multiple annotators on articles, so we were unable to compute inter-annotator agreement. This evaluation dataset of annotations is available at [DOI pending].

This collaborative, grounded theory, approach to annotating authorship is motivated by our ultimate goal of serving researchers studying journalism, communication, and beyond in a variety of linguistic and cultural contexts.  It is also informed by feminist data collection practices of engaging with multiple perspectives and being transparent about decision-making related to data processing \cite{suresh_towards_2022, dignazio_data_2020}. By opening a dialogue with researchers through the annotation process, we can better understand the phenomena they would like to capture when studying authorship and be alerted to potential cultural biases and shortcomings of existing solutions. For example, we became aware of the need to capture authorship via usernames on blogging platforms, such as LiveJournal, which remains important to scholars studying Russia \cite{wright_linux_2019}, and the practice of embedding author names in images, which could not be marked in LabelStudio. Some annotators were more engaged in the feedback process than others, which may also introduce bias into our evaluation dataset. Moreover, understanding the needs of researchers with regard to author extraction will be an ongoing process as journalism on the web continues to change, so there will always be room for improving the evaluation dataset. Grounded theory includes planning for continual improvement and evolution during the data collection and analysis process \cite{blandford_qualitative_2016}, which we embrace as part of iterative development of research tools. Despite limited resources, incorporating dialogue with annotators and qualitative data analysis led to a better understanding of the author extraction task and illuminated potential shortcomings of our tools and data. 

\subsection{Selection of Libraries}
After creating the dataset, we leveraged our own experiences and surveyed existing literature and GitHub statistics to identify some often used libraries that include author extraction in their feature sets (see  table  \ref{table:version_license}). Many not only extracted authors but also other fields such as published date, title text, top images, etc. Reviewing source code and documentation revealed that these libraries use a multitude of approaches including heuristic based rules (such as regex matching), ML models (such as neural networks), and combination of the two to accomplish their information extraction from web content. The tools evaluated included: 

\subsubsection{Newspaper3k}
Newspaper3k is an open source heuristic-based tool for extracting information from web articles. To extract author information, it first checks for html tags containing names by using regex pattern matching and checks for html tags containing some of the popular class names such as “author”, “byline”, and “dc. creator”.

\subsubsection{Trafilatura}
Trafilatura mainly uses python-readability\footnote{\url{https://github.com/buriy/python-readability}} and jusText\footnote{\url{https://github.com/miso-belica/jusText}}. Trafilatura’s extraction algorithm is based on a cascade of rule based filters and content heuristics \cite{barbaresi_trafilatura_2021}.

\subsubsection{news-please}
News-please is an open source python library that combines the power of multiple state-of-the-art libraries and tools, such as scrapy\footnote{\url{https://scrapy.org/}}, Newspaper3k\footnote{\url{https://github.com/codelucas/newspaper}}, and readability\footnote{\url{https://github.com/buriy/python-readability}} to extract information from online articles. It also has the capability to recursively go to each and every website provided in a news article to extract structured news content (known as spidering) \cite{hamborg_news-please_2017}.

\subsubsection{Go-Readability}
Go-Readability\footnote{\url{ttps://github.com/go-shiori/go-readability}}, a library in the Go programming language, is designed to identify the primary legible content and metadata within an HTML page. It operates by eliminating extraneous elements such as buttons, advertisements, background images, scripts, and more. The development of this package has been inspired from Mozilla's Readability.js\footnote{\url{https://github.com/mozilla/readability}}.

\subsubsection{ExtractNet}
ExtractNet is a machine learning based python library, mainly influenced by dragnet\footnote{\url{https://github.com/dragnet-org/dragnet/}}, that extracts information from web articles. It was built to avoid problems from the heuristic based libraries which have the tendency to rely on the tag property to extract author names \cite{peters_content_2013}.

\begin{table}[t]
\centering
\begin{tabular}{|p{0.25\linewidth}|p{0.65\linewidth}|}\hline
    \textbf{Language} & \textbf{NER model} \\ \hline
    English & dslim/bert-base-NER\\ 
    French & Jean-Baptiste/camembert-ner\\
    Hindi, & ai4bharat/indic-bert \\
    Spanish &  mrm8488/bert-spanish-cased-finetuned-ner \\
    Danish & Maltehb/danish-bert-botxo-ner-dane \\
    German & mschiesser/ner-bert-german \\
    Chinese & bert-base-chinese\\
    Greek & amichailidis/bert-base-greek-uncased-v1-finetuned-ner \\
    Urdu & themohal/ner\_bert\_urdu \\
    Russian & creat89/NER\_FEDA\_Ru\\ \hline
\end{tabular}
\caption[Caption for LOF]{Languages and the corresponding NER models from hugging-face\footnotemark}
\label{table:ner_model}
\end{table}

\footnotetext{https://huggingface.co/}
\subsubsection{Custom-NER (baseline)}
We also decided to establish a baseline against a naive custom approach. Building on advancements in transformer models, this custom pipeline identifies candidate authors via NER and then uses heuristics to pick the top candidates. We leverage open-source transformer models to extract entities from the website's text, picking the appropriate model and tokenizer based on content language (see  table  \ref{table:ner_model}). With this list of candidate named entities, we then compute the frequency of occurrence of each and filter down to the fewest occurring as potential candidates for authorship. This approach builds on the observation that author names are unlikely to be mentioned repeatedly within a single article. To make sure the recall of the approach is high, we chose the lowest three candidates as the authors of the news article. 

\begin{table}[t]
\centering
\begin{tabular}{|p{0.21\linewidth}|p{0.2\linewidth}|p{0.4\linewidth}|}\hline
    \textbf{Tool} & \textbf{version} & \textbf{License}\\ \hline 
    Newspaper3k & 0.2.8 & Apache License\\ \hline
    Trafilatura &  1.6.1 &  GNU General Public License v3.0 \\ \hline
    news-please & 1.5.33 & Apache License 2.0 \\ \hline
    Go-readability & 0.1.1  & MIT License  \\ \hline
    ExtractNet & 2.0.7 & MIT License  \\ \hline
\end{tabular}
\caption[Caption for LOF]{Version and the License of usage of the tools evaluated}
\label{table:version_license}
\end{table}

\subsection{Criteria for Evaluation}
Automated evaluation of the accuracy of author extraction is more nuanced than it might seem at first glance. Traditional Named Entity Recognition evaluation schemes relying on character or token offsets (beginning-inside-outside) cannot be used, since some extractors will pull correct author information from HTML metadata, which won't match the in-text annotation. Encoding human judgements of “correctness” is also challenging, particularly in relation to inclusion of punctuation, variations in capitalization, etc. F-measures used in previous evaluations encode  exact-match policies, which may be overly strict, since author names will likely require post-processing to account for expected variation across publications. To embrace this nuance we decided to move past binary judgements and focus on metrics that quantify the degree of alignment between the gold standard (manually created annotations) and results from each library. This approach helped us identify whether the tool not only identified the correct names, but also on how similar they are to the gold standard entities.  This specifically focused on normalized edit distance and ROUGE scores.

Edit distance or Levenshtein distance is a metric used to determine how close two strings are to each other.  It imposes a cost depending upon the steps needed to convert one string to another by operations such as addition, deletion or replacement of a character by assigning 1, 1 and 2 points respectively for the operations. Hence a lower score means that the strings are better aligned with each other. However, changing a single character results in the same edit distance regardless of the string length. In order to better capture the ratio of correct characters, we normalize edit distance by the maximum length between the extracted and the gold standard  strings \cite{de_la_higuera_contextual_2008}. Normalized edit distance is helpful to reduce the penalization for longer length strings and make comparisons across languages with different average word length on the same scale. Using this measure helped us focus on understanding how far from correct a response was, capturing something closer to a human judgment.

Another well known metric used in evaluating the efficacy of NLP models is ROUGE \cite{lin_rouge_2004}. It is used to evaluate the performance of NLP in the task of abstractive and extractive text summarization.  Defined broadly, ROUGE is used to compare subsequences  between two strings. In our case, we compare character subsequences between candidate author names and the gold standard reference. To make sure the comparison is fair, we first strip off the extra characters and convert every character to its lowercase form. Later, we sort the author names on the basis of their lexical order. The ROUGE metric can be modified to compare subsequences of different lengths, also called character n-grams, where n is the number of adjacent characters used in counting matches. Small ROUGE scores like ROUGE-1 would not take into account the order of letters in the strings that are getting compared. Hence, it is preferable to use larger ROUGE scores where longer substrings must match for a candidate string to get a high ROUGE score. In addition, the ROUGE-L score determines the longest subsequence of character tokens  overlapping between the two strings. Using the ROUGE metric offered an alternative take on overlaps between the gold standard and the answers produced by each library, again giving us a more human-like blurry concept of “correctness” than a binary match would.

\begin{figure*}[t]
\centering
\includegraphics[width=0.8\textwidth]{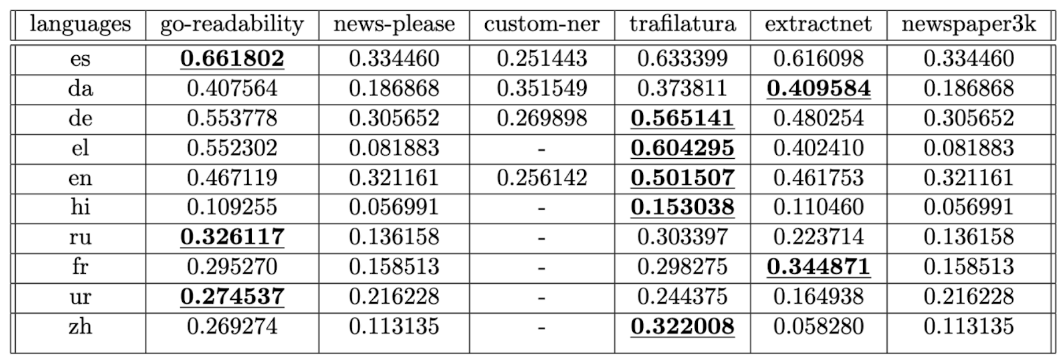} 
\caption{Rouge-1 scores for each library in each language. Higher is better. Dashes indicate instances of no character overlap.}
\label{table:rouge1}
\end{figure*}

\begin{figure*}[t]
\centering
\includegraphics[width=0.8\textwidth]{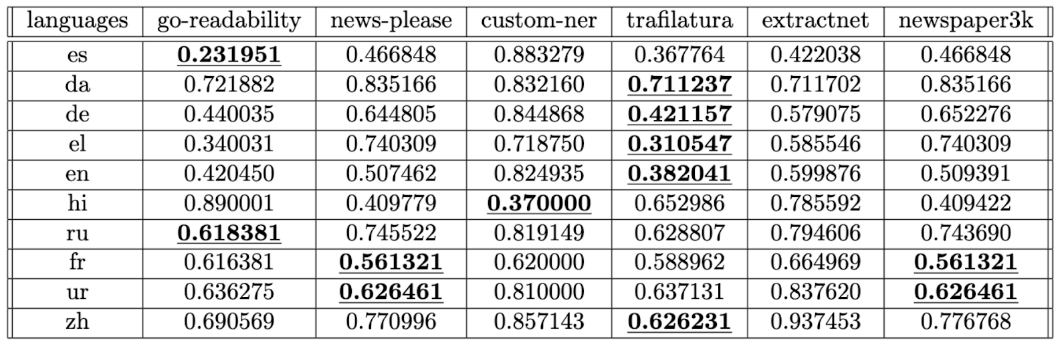} 
\caption{Normalized edit distance scores for each library in each language. Lower is better.}
\label{table:normalized_edit_dist}
\end{figure*}

\begin{figure*}[t]
\centering
\includegraphics[width=0.8\textwidth]{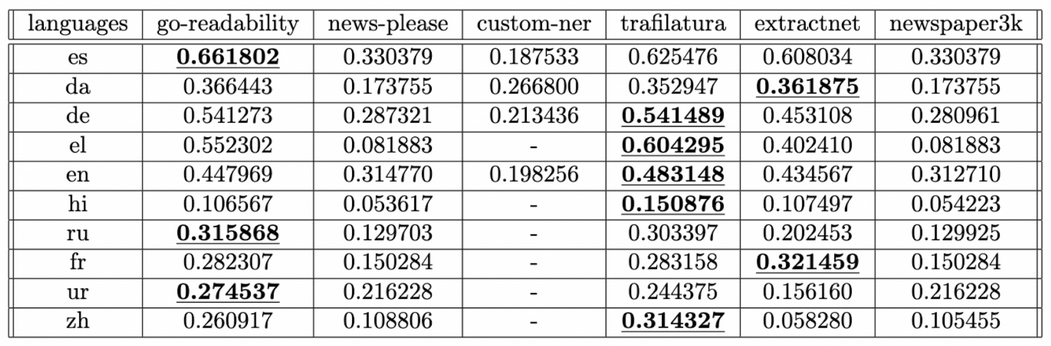} 
\caption{Rouge-L scores for each library in each language. Higher is better. Dashes indicate instances of no character overlap.}
\label{table:rougeL}
\end{figure*}

\section{Results}
Figures  \ref{table:rouge1}, \ref{table:normalized_edit_dist} and \ref{table:rougeL} contain the metrics for each language and tool pair, determined by averaging the tool's scores for each document in that language's evaluation set. Each table includes a row for each language within the corpus, and a column for each library used. Each cell is the computed score for that metric. The best score for each language is highlighted in bold and underlined. Our evaluation code is available on GitHub\footnote{anonymizedURL}.

\begin{figure*}[t]
\centering
\includegraphics[width=1\textwidth]{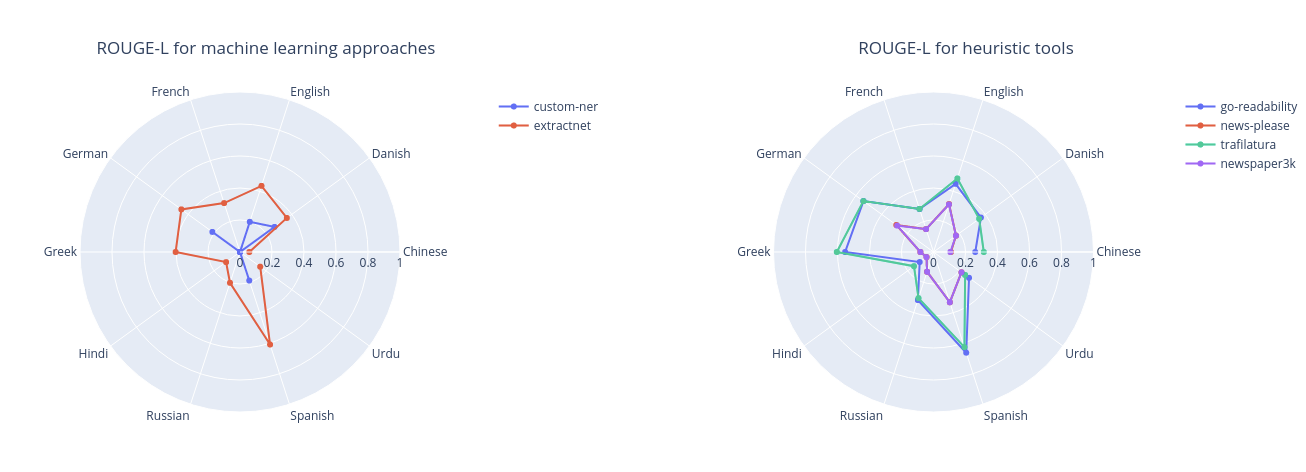} 
\caption{Radar plot of ROUGE-L scores}
\label{fig:radarplot-rougeL}
\end{figure*}

\begin{figure*}[t]
\centering
\includegraphics[width=1\textwidth]{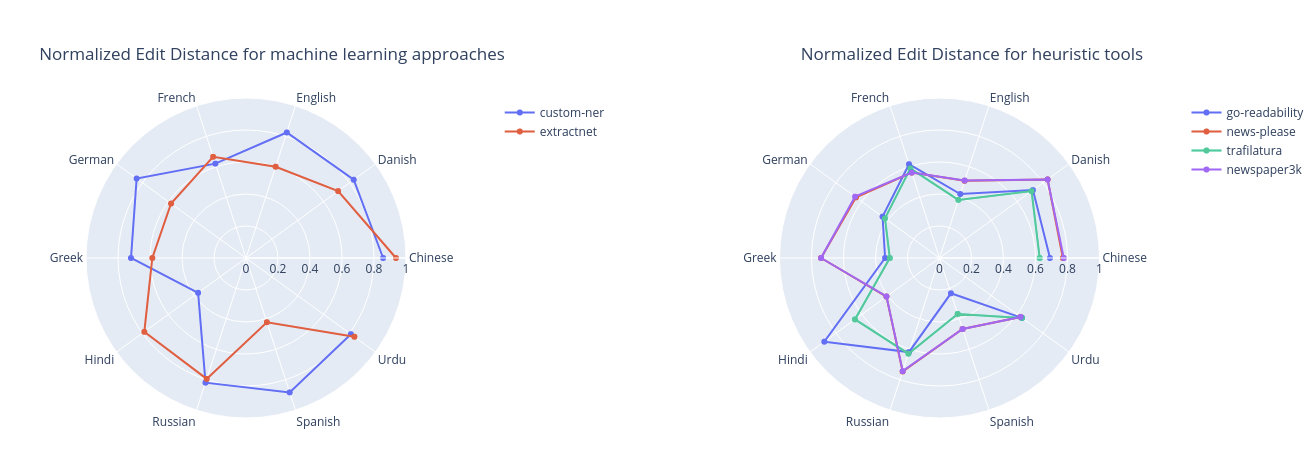} 
\caption{Radar plot of normalized edit distances}
\label{fig:radarplot-normalized-edit}
\end{figure*}

Overall the results reveal no obvious single library that performs at a high level across all languages (see figures  \ref{fig:radarplot-rougeL} and \ref{fig:radarplot-normalized-edit}). For news-articles in languages German, Greek, English and Chinese, \textbf{Trafilatura} performs the best on both ROUGE and normalized edit distance metric. For other news articles in languages such as Spanish, Russian and Urdu \textbf{Go-readability} performs the best according to ROUGE scores. However, for Urdu news articles, according to normalized edit distance, \textbf{news-please} performs the best. News-please and Newspaper3k perform  similarly, unsurprising given that the former depends on the latter. No tool currently performs sufficiently good enough for news articles in Hindi. The best tool as per the metric is Trafilatura with a low score of 0.15 in ROUGE-L score and our custom named entity recognizer (NER) with 0.37 as normalized edit distance score. 

Evaluating author retrieval accuracy revealed Trafilatura to be the most effective tool, while custom NER demonstrated the weakest performance. This disparity can be attributed to custom NER's inability to determine the intended number of authors per article. These findings underscore the limitations of naive named entity recognition approaches in tackling this intricate task. Consequently, further investigation into exploration capable of leveraging web article’s DOM nodes and markup modeling (akin to language modeling)  is needed to research their potential in author retrieval. We elaborate on promising research directions in the future work section.

\section{Discussion}
This evaluation of author extraction libraries offers a set of results to guide researchers in choosing the most appropriate for their specific projects. While some libraries outperform others, overall it is clear that the performance of the libraries varies widely based on language, and no single library offers a satisfactory solution across the board. Users of these libraries are advised to extend the validation process presented here and use contextually appropriate sample articles and languages for that purpose.

The analysis has a few limitations worth noting. First, the sample size is fairly small (100 articles per language) and limited to the specific language our team of annotators spoke. Second, this is a point-in-time snapshot of the performance of specific versions of the libraries discussed; we anticipate the results for any major revisions would vary. Despite these limitations, validation efforts such as this are critical for supporting robust analysis that relies on authors extracted via any of the libraries evaluated.

\subsection{Socio-cultural Contexts}
Contextually, it is important to recognize that the structure of news online is both technologically and socially determined. Modern web pages look different than those from just a few years ago, but more importantly they are often structured differently. The adoption and use of HTML \textless meta\textgreater tags in news sites ebbs and flows based on social media use. Search engine guidance on the use of tag attributes (such as `rel=``author"' or `class=``author"') regularly changes. The constant evolution of web coding norms, and related use in social media sites and search engines, suggests that heuristics in libraries might not age well. Newspaper3k hasn't had a release since 2018, while Trafilatura has seen regular updates since its release in 2019.

In addition, the sample dataset and prior experience has shown us that the content management systems hosting and rendering news sites have very regional patterns of use.  For instance, Wordpress might be a standard in one country for hosting news sites, but totally unused in another. These content management systems often hold the rules about how an author’s name is rendered, meaning that norms of hosting platform use in one country might end up meaning that a particular library functions well, or not well, in that context. This is one of the motivations for creating an evaluation benchmark with global coverage (beyond the obvious desire to avoid cultural bias).

\subsection{Ethical Concerns}
At a small scale the task of author extraction doesn’t expose any particular ethical concerns. The computational task relies on the extraction of already published information, moving it from unstructured to structured form. That said, any methods that rely on content could perform poorly cross-culturally and/or cross-linguistically, a sensitivity that could significantly impact research results in global settings. For instance, the authors have anecdotal evidence that existing solutions perform better on English-language corpora of articles from US-based publishers.

However, we must consider broader scaled-up usage of author extraction technologies within the larger context of growing violence against journalists and newspapers across the world \cite{international_federation_of_journalists_one_2023}. This is particularly true of journalists of minoritized identities \cite{posetti_online_2020}. If publicly available large-scale corpora of online news integrate authorship, they risk exposing journalists who write about contentious issues to bad actors looking to silence them through online or in-person threats. Consider the hypothetical of a bad-actor looking to silence reporters writing about transgender rights in the United States. If they were able to search and download a corpus of published news articles including positive framing of transgender rights that included author names, they could mobilize a harassment campaign at scale against the authors of those articles.

Any researcher or software developer deploying author extraction at scale must consider the potential harms of its use, particularly the exposure for discovery of patterns of reporting by authors who write about contested issues.

\section{Future Work}
Our existing methods for author extraction from online news articles, including heuristic-based systems and basic NLP techniques like named entity recognition, have proven inadequate. Moving forward, promising avenues lie in leveraging the website's structure (DOM nodes) for classification by creating a comprehensive repository of DOM node sets and training a classifier to identify nodes likely containing author information \cite{lin_freedom_2020,zhou_learning_2022}. Furthermore, recent advancements in transformer-based neural networks, particularly masked language modeling techniques, open up possibilities for representation learning within semi-structured data like HTML \cite{deng_dom-lm_2022}. By fine-tuning these pre-trained models, we can effectively extract author-containing nodes, but to the best of our knowledge this has not been tried specifically for the author extraction task yet. Additionally, the emergence of open-source LLMs like Llama \cite{touvron_llama_2023} and Mistral \cite{jiang_mistral_2023} presents another potential avenue. We can experiment with carefully crafted prompts incorporating the entire visual and textual data of a website to guide the LLM toward accurate author retrieval. Leveraging more modern language models holds significant promise for tackling the challenge of automatic author extraction in the vast landscape of online news articles.

\section{Conclusion}
Effective author extraction from online news articles has great potential to open many new avenues of research in journalism and communication studies as well as applied NLP and will help us better understand media ecosystems in the long-term. However, before any author extraction tool can be used in a news research archive, we must understand its behavior across linguistic, geographical and cultural contexts and ensure researchers trust the output. To this end, we relied on participatory research practices to build an author extraction evaluation data set in collaboration with users of an online news archive. 

We used this dataset to evaluate popular out-of-the-box author extraction libraries and a novel baseline method relying on state-of-the-art NER models. Despite the prevalence of out-of-the-box author extraction tools and a rich history of work on the problem in web and data mining research, we find that no single approach effectively captures authorship across all languages, although Trafilatura performs best on 5 out of the 10 languages in our dataset. Our analysis reveals challenges both from socio-cultural perspectives, including linguistic variation and different journalistic conventions, and on account of socio-technical systems, such as changing conventions around HTML metadata, popularity of various content management systems and the maintenance of author extraction libraries. 

We hope that this paper gives fellow researchers a better picture of the current state of author extraction and motivates future work collecting evaluation data across socio-cultural contexts and employing transformer-based machine learning techniques to create more robust, flexible author extractors. We also encourage engagement in understanding the downstream applications of author extraction in fields such as computational journalism, media communications and journalism studies,  since this can lead to deeper understanding of the socio-cultural factors that make this problem particularly challenging. 


\bibliography{author_extraction}

\section{Paper Checklist}

\begin{enumerate}

\item For most authors...
\begin{enumerate}
    \item  Would answering this research question advance science without violating social contracts, such as violating privacy norms, perpetuating unfair profiling, exacerbating the socio-economic divide, or implying disrespect to societies or cultures?
    \answerYes{Yes}
  \item Do your main claims in the abstract and introduction accurately reflect the paper's contributions and scope?
    \answerYes{Yes}
   \item Do you clarify how the proposed methodological approach is appropriate for the claims made? 
    \answerYes{Yes}
   \item Do you clarify what are possible artifacts in the data used, given population-specific distributions?
    \answerYes{Yes}
  \item Did you describe the limitations of your work?
    \answerYes{Yes}
  \item Did you discuss any potential negative societal impacts of your work?
    \answerYes{Yes}
      \item Did you discuss any potential misuse of your work?
    \answerYes{Yes}
    \item Did you describe steps taken to prevent or mitigate potential negative outcomes of the research, such as data and model documentation, data anonymization, responsible release, access control, and the reproducibility of findings?
    \answerYes{Yes}
  \item Have you read the ethics review guidelines and ensured that your paper conforms to them?
    \answerYes{Yes}
\end{enumerate}

\item Additionally, if your study involves hypotheses testing...
\begin{enumerate}
  \item Did you clearly state the assumptions underlying all theoretical results?
    \answerNA{NA}
  \item Have you provided justifications for all theoretical results?
    \answerNA{NA}
  \item Did you discuss competing hypotheses or theories that might challenge or complement your theoretical results?
    \answerNA{NA}
  \item Have you considered alternative mechanisms or explanations that might account for the same outcomes observed in your study?
    \answerNA{NA}
  \item Did you address potential biases or limitations in your theoretical framework?
    \answerNA{NA}
  \item Have you related your theoretical results to the existing literature in social science?
    \answerNA{NA}
  \item Did you discuss the implications of your theoretical results for policy, practice, or further research in the social science domain?
    \answerNA{NA}
\end{enumerate}

\item Additionally, if you are including theoretical proofs...
\begin{enumerate}
  \item Did you state the full set of assumptions of all theoretical results?
    \answerNA{NA}
	\item Did you include complete proofs of all theoretical results?
    \answerNA{NA}
\end{enumerate}

\item Additionally, if you ran machine learning experiments...
\begin{enumerate}
  \item Did you include the code, data, and instructions needed to reproduce the main experimental results (either in the supplemental material or as a URL)?
    \answerYes{Yes}
  \item Did you specify all the training details (e.g., data splits, hyperparameters, how they were chosen)?
    \answerNA{NA}
     \item Did you report error bars (e.g., with respect to the random seed after running experiments multiple times)?
    \answerNA{NA}
	\item Did you include the total amount of compute and the type of resources used (e.g., type of GPUs, internal cluster, or cloud provider)?
    \answerNo{No, because we were evaluating existing libraries and this didn't require GPUs or large compute resources.}
     \item Do you justify how the proposed evaluation is sufficient and appropriate to the claims made? 
    \answerYes{Yes}
     \item Do you discuss what is ``the cost`` of misclassification and fault (in)tolerance?
    \answerYes{Yes} 
  
\end{enumerate}

\item Additionally, if you are using existing assets (e.g., code, data, models) or curating/releasing new assets, \textbf{without compromising anonymity}...
\begin{enumerate}
  \item If your work uses existing assets, did you cite the creators?
    \answerYes{Yes}
  \item Did you mention the license of the assets?
    \answerYes{Yes}
  \item Did you include any new assets in the supplemental material or as a URL?
    \answerYes{Yes}
  \item Did you discuss whether and how consent was obtained from people whose data you're using/curating?
    \answerYes{Yes}
  \item Did you discuss whether the data you are using/curating contains personally identifiable information or offensive content?
    \answerYes{Yes}
\item If you are curating or releasing new datasets, did you discuss how you intend to make your datasets FAIR?
\answerYes{Yes}
\item If you are curating or releasing new datasets, did you create a Datasheet for the Dataset? 
\answerNo{No, but we will do so when the dataset is released publicly upon publication.}
\end{enumerate}

\item Additionally, if you used crowdsourcing or conducted research with human subjects, \textbf{without compromising anonymity}...
\begin{enumerate}
  \item Did you include the full text of instructions given to participants and screenshots?
    \answerNo{No, but annotation guidelines will be released publicly with the dataset.}
  \item Did you describe any potential participant risks, with mentions of Institutional Review Board (IRB) approvals?
    \answerNA{NA}
  \item Did you include the estimated hourly wage paid to participants and the total amount spent on participant compensation?
    \answerNA{NA}
   \item Did you discuss how data is stored, shared, and deidentified?
   \answerNA{NA}
\end{enumerate}

\end{enumerate}

\end{document}